\documentstyle[12pt]{article}
\begin{document}
\vskip 0.1in
\centerline{\Large\bf On the viability of the Palatini} 
\centerline{\Large\bf form of 1/R gravity}
\vskip .7in
\centerline{Dan N. Vollick}
\centerline{Department of Physics and Astronomy}
\centerline{and}
\centerline{Department of Mathematics and Statistics}
\centerline{Okanagan University College}
\centerline{3333 College Way}
\centerline{Kelowna, B.C.}
\centerline{V1V 1V7}
\vskip .9in
\centerline{\bf\large Abstract}
\vskip 0.5in
Recently Flanagan \cite{Fl1} has argued that the Palatini form 
of 1/R gravity is ruled out by experiments such as electron-electron 
scattering. His argument involves adding minimally coupled fermions in the
Jordan frame and transforming to the Einstein frame. This produces additional
terms that are ruled out experimentally.
  
Here I argue that this conclusion is false. It is well known that 
conformally related theories are mathematically equivalent but not
physically equivalent.
As discussed by Magnano and Sokolowski \cite{Ma1} one must decide, in the
vacuum theory,
which fame is the physical frame and add the minimally coupled Lagrangian in this
frame. If this procedure is followed the resulting theory is not ruled out
experimentally.
The discussions in this paper also show that the equivalence between  
the generalized gravitational theories and scalar tensor theories 
discussed by Flanagan \cite{Fl2} is only mathematical, not physical.
\newpage
Recently there have been several papers that attempt to explain the observed
cosmological acceleration by including a $1/R$ term in the gravitational
Lagrangian. Capozziello et al. \cite{Cap1} and Carroll et al. 
\cite{Ca1} used
a purely metric variation to obtain the field equations. This approach
was shown to be in conflict with solar system observations by Chiba \cite{Ch1}.
Noriji and Odintsov \cite{No1,No2} subsequently showed that this 
problem may be cured
by adding higher positive powers of the Ricci scalar to the action.
I obtained the field equations by using a Palatini variation and
showed that the resulting field equations can explain the cosmological acceleration
\cite{Vo1}.
  
Recently, Flanagan \cite{Fl1} has argued that the Palatini form of the theory
is in conflict with experiments such as electron-electron scattering.
In his paper he writes the theory in a mathematically equivalent form by
transforming the gravitational action plus a minimally coupled matter action
to the Einstein frame (see also \cite{Fl2}).
This introduces a new scalar field without a kinetic
term. Integrating out the field produces corrections to the standard model
that are experimentally ruled out.

The key problem that invalidates this
argument is the failure to properly identify the physical 
frame and to add the minimally coupled matter Lagrangian in that frame.
A frame will be called the physical frame if the variables that appear
in the formulation of the theory in that frame are measurable and satisfy
all of the requirements of a classical field theory \cite{Ma1}. 
Since the metrics in
conformally related frames are different it is only possible for one of them
to describe the metrical structure of our spacetime manifold. This structure
can be determined experimentally and will determine which of the
metrics is the physical metric. A simple example, given by Faraoni
et al. \cite{Fa1}, is a k=0 FRW universe. The metric, in conformal coordinates,
is given by
\begin{equation}
ds^2=a(\eta)^2[-d\eta^2+dx^2+dy^2+dz^2]\; .
\end{equation}
This metric is conformally equivalent to the Minkowski metric but the Minkowski
metric is not physical as it does not describe the metrical properties of
the FRW universe. 
   
The physical inequivalence of the Jordan and Einstein frames
in scalar tensor theories and in nonlinear gravitational theories, i.e. theories
with a Lagrangian that is an arbitrary function of the Ricci scalar, has
been shown by many authors. All of these arguments have been formulated within
the usual purely metric variation approach.
Bellucci et al. \cite{Be1} showed that the Einstein
and Jordan frame descriptions of the interaction of scalar gravitational
waves with a detector in Brans-Dicke theory differ. Faraoni and Gunzig 
\cite{Fa2} showed
that gravitational waves in Brans-Dicke theory violate the weak energy condition
in the Jordan frame but not in the Einstein frame. In scalar tensor theories
of gravity the energy density of the gravitational scalar field can violate
the weak energy condition in the Jordan frame, but not in the Einstein frame
\cite{Ma1,Fa1}. Difficulties have also been
shown \cite{Sa1,Ko1,Fu1,Ni1} to occur in quantizing scalar field 
fluctuations, in
the linear approximation, in the Jordan frame but not in the Einstein frame.
This indicates that quantization and conformal transformations do not commute.
Magnano and Sokolowski \cite{Ma1} showed that the Einstein frame 
is the only physical
frame in nonlinear gravitational theories. All other frame, including the
Jordan frame, are not physical. They showed that violations of the dominant
energy condition occur in all frames except the Einstein frame. These examples
should suffice to show that the Einstein and Jordan frames are not physically
equivalent. In fact, these examples indicate that for theories derived from
a purely metric variation it is the Einstein frame
that is physical. In general conformally related theories are mathematically
equivalent in the sense that there solution spaces are isomorphic, but they
are not physically equivalent. This will, of course, also apply to the equivalence
between the gravitational and scalar tensor theories discussed by Flanagan
in \cite{Fl2}.
  
Now consider the inclusion of matter by adding a minimally coupled matter
Lagrangian to the gravitational Lagrangian. In which frame do we add
this Lagrangian? It is obvious that we get different theories if we add it
in the Jordan frame or in the Einstein frame. We must first decide, in the
vacuum theory, which of the frames is physical (i.e. which metric is physical)
and add the minimally coupled Lagrangian in this frame (see Magnano and Sokolowski
\cite{Ma1}). This ensures that particle masses are constant in the physical
frame. Thus, one cannot add a minimally coupled matter Lagrangian in one
frame and then transform to another frame and conclude that the theory is
ruled out because of the nonminimally coupled terms in the new frame. 
However, this is exactly what Flanagan has done to argue that the 
Palatini form of 1/R gravity is ruled out. If
minimally coupled matter is added in a frame this must be the
physical frame and all predictions of the theory must be calculated in this
frame.

Now consider Flanagan's argument on the non-viability of the Palatini approach
to 1/R gravity in more detail. The action is
\begin{equation}
S[\bar{g}_{\mu\nu},H^{\lambda}_{\mu\nu},\psi_m]=\frac{1}{2\kappa^2}\int d^4x\sqrt{-\bar{g}}
f(\hat{R})+S_m[\bar{g}_{\mu\nu},\psi_m]\; ,
\label{action1}
\end{equation}
where $\bar{g}_{\mu\nu}$ is the Jordan frame metric, $\psi_m$ are the 
matter fields, $H^{\lambda}_{\mu\nu}$ is defined by
\begin{equation}
\hat{\nabla}_{\mu}v^{\lambda}-\bar{\nabla}_{\mu}v^{\lambda}=H^{\lambda}_{\mu\nu}v^{\nu}
\end{equation}
for any vector $v^{\lambda}$, $\hat{\nabla}_{\nu}$ is a symmetric connection,
$\bar{\nabla}_{\mu}$ is the connection associated with the metric $\bar{g}_{\mu\nu}$,
$\hat{R}=\bar{g}^{\mu\nu}\hat{R}_{\mu\nu}$ and
\begin{equation}
\hat{R}_{\mu\nu}=\bar{R}_{\mu\nu}+\bar{\nabla}_{\lambda}H^{\lambda}_{\mu\nu}
-\bar{\nabla}_{\mu}H^{\lambda}_{\lambda\nu}+H^{\lambda}_{\lambda\sigma}
H^{\sigma}_{\mu\nu}-H^{\lambda}_{\mu\sigma}H^{\sigma}_{\lambda\nu}.
\end{equation}
This action is then shown to be mathematically equivalent to the action
\begin{equation}
\tilde{S}[g_{\mu\nu},\Phi,\psi_m]=\int d^4x\sqrt{-g}\left[\frac{R}{2\kappa^2}-V(\Phi)
\right]+S_m[e^{2\alpha(\Phi)}g_{\mu\nu},\psi_m],
\label{action2}
\end{equation}
where 
\begin{equation}
V(\Phi)=\frac{\Phi f^{'}(\Phi)-f(\Phi)}{2\kappa^2f^{'}(\Phi)^2}
\end{equation}
and
\begin{equation}
g_{\mu\nu}=e^{-2\alpha(\Phi)}\bar{g}_{\mu\nu}
\end{equation}
is the Einstein frame metric. 
   
For 1/R gravity the function $f(R)$ is taken to be
\begin{equation}
f(R)=R-\frac{\mu^4}{R}\; .
\end{equation}
Flanagan then takes the matter
action to be the action for free electrons:
\begin{equation}
S_m[\bar{g}_{\mu\nu},\Psi_e]=\int d^4x\bar{\Psi}_e[i\bar{\gamma}^{\mu}
\bar{\nabla}_{\mu}-m_e]\Psi_e\; .
\label{mat}
\end{equation}
In the Einstein frame the action becomes
\begin{equation}
\tilde{S}[g_{\mu\nu},\Phi,\Psi_e]=\int d^4x\sqrt{-g}\left[\frac{R}{2\kappa^2}-
V(\Phi)+ie^{3\alpha(\Phi)}\bar{\Psi}_e\gamma^{\mu}\nabla_{\mu}\Psi_e-e^{4\alpha(\Phi)}m_e
\bar{\Psi}_e\Psi_e\right]\; .
\end{equation}
Finally, he integrates out $\Phi$ by solving its equations of motion (approximately)
and substituting back into the action to get
\begin{equation}
\begin{array}{ll}
\tilde{S}[g_{\mu\nu},\Psi_e]=\int d^4x\sqrt{-g}[\frac{R}{2\tilde{\kappa}^2}-\Lambda
+i\bar{\Psi}_e\gamma^{\mu}\nabla_{\mu}\Psi_e-m_e\bar{\Psi}\Psi_e-\frac{3\sqrt{3}}{16
m_*^4}(i\bar{\Psi}_e\gamma^{\mu}\nabla_{\mu}\Psi_e)^2\\
\;\\
-\frac{1}{\sqrt{3}}\frac{m_e^2}{m_*^4}(\bar{\Psi}_e\Psi_e)^2+\sqrt{\frac{3}{4}}
\frac{m_e}{m_*^4}(i\bar{\Psi}_e\gamma^{\mu}\nabla_{\mu}\Psi_e)(\bar{\Psi}_e\Psi_e)+...
]\; ,\\
\end{array}
\end{equation}
where $\tilde{\kappa}=\sqrt{4/3}\kappa$, $m_*=\sqrt{\mu/\kappa}$ and $\Lambda=
\mu^2/(\sqrt{3}\kappa^2)$.
The last three terms in the above action correspond to corrections in the
standard model, characterized by the mass scale $m_*\sim 10^{-3}$eV. Corrections
with such a small mass scale are ruled out experimentally (e.g. in electron-electron
scattering). Thus, he concludes that the original theory, given by the action
(\ref{action1}) is ruled out.
  
The problem with argument is that the minimally coupled Lagrangian is added
to the original Jordan frame Lagrangian (\ref{action1}), so that this must
be the physical frame. Thus, all calculations must be done in this frame
and it is easy to see from (\ref{action1}) and from (\ref{mat}) that no corrections
to the standard model arise that are experimentally ruled out. As discussed
earlier, one cannot add a minimally coupled matter Lagrangian in one
frame and then transform to another frame and conclude that the theory is
ruled out because of the nonminimally coupled terms in the new frame.
If one wants to take the Einstein metric as physical the minimally coupled
Lagrangian has to be added in that frame. In this case the matter Lagrangian
is independent of $\Phi$ and the $\Phi$ equation of motion is $V^{'}(\Phi)=0$.
Thus, $\Phi$ is a constant and one obtains Einstein's theory with a 
cosmological constant and a minimally coupled Dirac field. As in the original
version of the theory the value of $\mu$ can be chosen to give a cosmological
constant that can produce the observed cosmological acceleration. This 
shows that the theory is therefore viable in either conformal frame.
   
\section*{Acknowledgements}
I would like to thank Valerio Faraoni for helpful discussions on the 
physical inequivalence of conformally related frames. 


\begin{thebibliography}{1}
\bibitem{Fl1}
E.E. Flanagan, astro-ph/0308111
\bibitem{Ma1}
G. Magnano and L.M. Sokolowski, Phys. Rev. D50, 5039 (1999), 
gr-qc/9312008
\bibitem{Fl2}
E.E. Flanagan, Class. Quant. Grav. 21, 417 (2003), gr-qc/0309015
\bibitem{Cap1}
Capozziello, S. Carloni and A. Troisi, astro-ph/0303041
\bibitem{Ca1}
S.M. Carroll, V. Duvvuri, M. Trodden and M.S. Turner, astro-ph/0306438
\bibitem{Ch1}
T. Chiba, astro-ph/0307338
\bibitem{No1}
S. Nojiri and S.D. Odintsov, hep-th/0307288
\bibitem{No2}
S. Nojiri and S.D. Odintsov, hep-th/0310045
\bibitem{Vo1}
D.N. Vollick, Phys. Rev. D68, 063510 (2003), astro-ph/0306630
\bibitem{Fa1}
V. Faraoni, E. Gunzig, P. Nardone, Fundamentals of Cosmic Physics 20, 121
(1999), gr-qc/9811047
\bibitem{Be1}
S. Bellucci, V. Faraoni and D. Babusci, hep-th/0103180
\bibitem{Fa2}
V. Faraoni and E. Gunzig, Int. J. Theor. Phys. 38, 217 (1999), 
astro-ph/9910176
\bibitem{Sa1}
D. Salopek, J. Bond and J. Bardeen, Phys. Rev. D40, 1753 (1989)
\bibitem{Ko1}
E.W. Kolb, D. Salopek and M.S. Turner, Phys. Rev. D42, 3925 (1990)
\bibitem{Fu1}
Y. Fujii and T. Nishioka, Phys. Rev. D42, 361 (1990)
\bibitem{Ni1}
T. Nishioka and Y. Fujii, Phys. Rev. D45, 2140 (1992)
\end{thebibliography}
\end{document}